\documentclass[10pt]{article}
\usepackage{geometry}                
\usepackage{graphicx}
\usepackage{amssymb}
\usepackage{amsmath}
\usepackage{authblk}
\usepackage{bm}
\usepackage{cite}

\usepackage{epstopdf}
\usepackage{cite}
\begin{document}

\title{\bf Shaping van der Waals nanoribbons via torsional constraints: Scrolls, folds and supercoils}

\author{Alireza Shahabi$^a$, Hailong Wang$^b$ and Moneesh Upmanyu$^a$}
\affil{\small $^a$Group for Simulation and Theory of Atomic-scale Material Phenomena (stAMP), Department of Mechanical and Industry Engineering, Northeastern University, Boston, Massachusetts 02115, USA \\
\small $^b$Department of Materials Science and Engineering, University of Pennsylvania, Philadelphia, PA, 19104}

\date{}

\maketitle

{\bf Abstract: Interplay between structure and function in atomically thin crystalline nanoribbons is sensitive to their conformations yet the ability to prescribe them is a formidable challenge. Here, we report a novel paradigm for controlled nucleation and growth of scrolled and folded shapes in finite-length nanoribbons. All-atom computations on graphene nanoribbons (GNRs) and experiments on macroscale magnetic thin films reveal that decreasing the end distance of torsionally constrained ribbons below their contour length leads to formation of these shapes. The energy partitioning between twisted and bent shapes is modified in favor of these densely packed soft conformations due to the non-local van Der Waals interactions in these 2D crystals; they subvert the formation of supercoils that are seen in their natural counterparts such as DNA and filamentous proteins. The conformational phase diagram is in excellent agreement with theoretical predictions. The facile route can be readily extended for tailoring the soft conformations of crystalline nanoscale ribbons, and more general self-interacting filaments.}

\pagebreak
The field of atomically thin crystalline films continues to grow, both in terms of the amenable material systems and routes for processing and manipulating them.  Nanoribbons of these layered 2D materials exhibit superior functional properties that are sensitive to the ribbon growth direction~\cite{gnr:SonCohenLouie:2006}, the core and edge structure, and the interlayer interactions. The properties can be further tuned by the structure of the edges and the inherent coupling between the layers\cite{graph:BragaGalvaoBaughman:2004}.  For example, Archimedean scrolls of graphene exhibit tunable transport\cite{gnr:ShiGao:2009, gnr:MartinsGalvao:2010, gnr:LiLin:2012}, supercapacitance\cite{graph:ZengHaihui:2012}, and enhanced hydrogen storage\cite{gnr:BragaGalvao:2007}. Similarly, folds in graphene, or grafolds\cite{graph:KimZettl:2011}, can effect semiconducting-metallic transitions\cite{gnr:XieZhong:2009}, and increase the material strength\cite{graph:KimZettl:2011} by localizing strain accommodation within the ribbons.

The functional properties of these ribbons both influence, and are influenced by their conformations. In particular, the interplay between structure, geometry and conformation is sensitive to the extent of confinement, indicating the possibility of reversibly engineering their shapes by manipulating their end constraints.  Some of the well-known shapes include include twisted and helical ribbons, driven by changes in edge structure, chemistry and ribbon geometry~\cite{gnr:YuRuoff:2001, gnr:ChuvlinKhlobystov:2011, gnr:CranfordBuehler:2011, gnr:WangUpmanyu:2012, gnr:WangUpmanyuII:2012}. These conformations are topologically invariant and since the nature of atomic-scale interactions remains fundamentally unchanged, their effect on the properties is often limited. The ability to engineer conformations with topologies that enhance non-local, interlayer interactions - scrolls, folds, and knots - can  and in some cases novel properties, yet this remains a challenge due to the difficulties in manipulating the end constraints. 

\begin{figure}
\centering
\includegraphics[width=0.8\columnwidth]{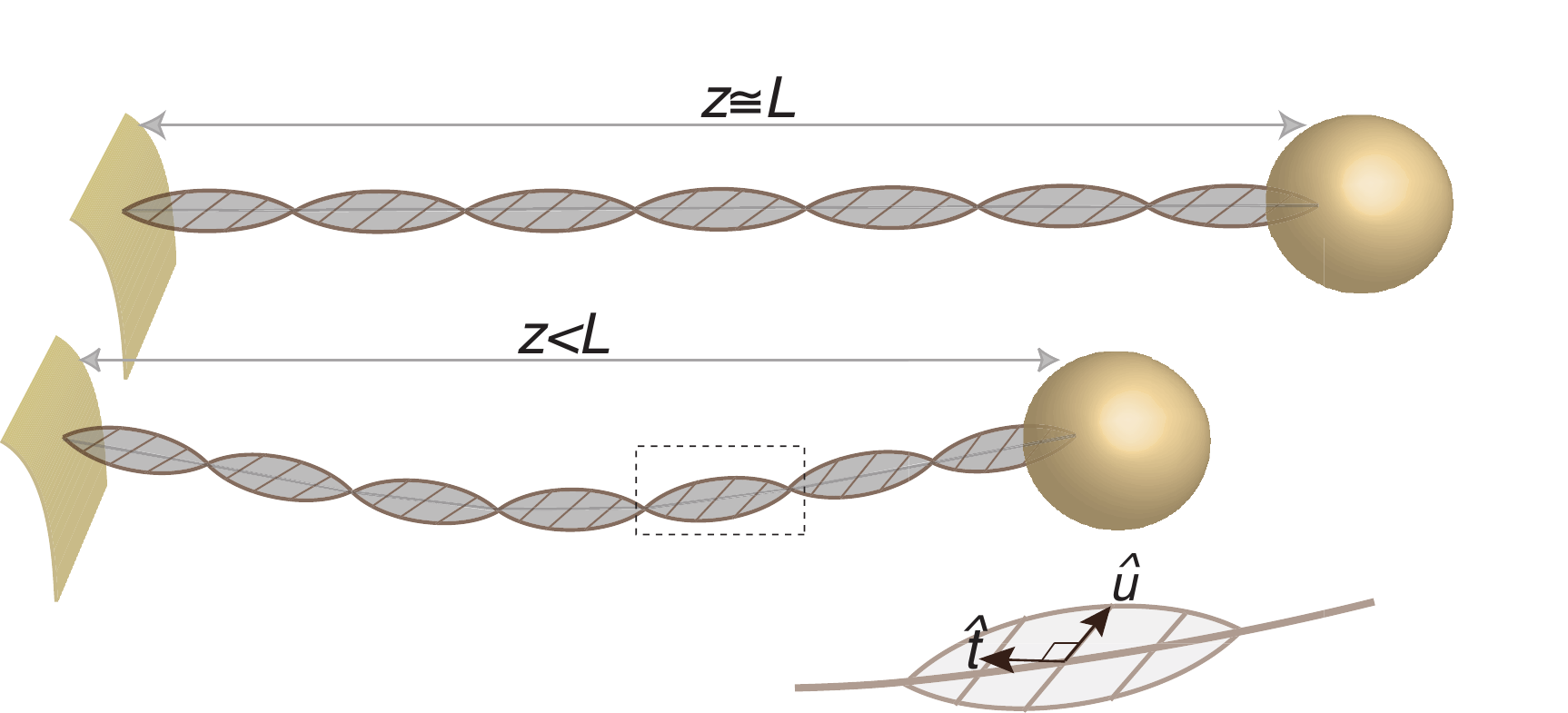}
\caption{\label{fig:fig1} (top) Schematic of a suspended supercoiled nanoribbon subject to a torsional constraint. The relative rotation between the two ends sets the degree of supercoiling $Lk$. (bottom) The strategy used to explore the  bent and twisted conformations for $z<L$. Magnified view of the ribbon (boxed) illustrating the ribbon and tangent vectors, $\hat{u}$ and $\hat{t}$ respectively, associated with the material frame used to describe the ribbon conformation.}
\end{figure}

In this article, we present a facile route to engineering topologically distinct soft conformations of nanoscale ribbons. Figure~\ref{fig:fig1} depicts the scenario schematically; a finite-length nanoribbon of width $w$ is torsionally constrained by rotating one end relative to the other and clamping the two ends. The end conditions take the form of a fixed degree of supercoiling $Lk$ and controlled end displacement $\lambda=z/L$.
The choice is motivated by the fact that, unlike the end couple (moment $M$ and tension $T$), 
the rigid loading variables $\lambda$ and $Lk$ are more accessible and can be easily manipulated~\cite{elastica:ThompsonChampneysI:1996}. For a ribbon so constrained, the partitioning of the initial twist (the Twist, $Tw$) into energetically favorable bent shapes (the Writhe, $Wr$) follows from the well-known C{\u a}lug{\u a}reanu-White-Fuller theorem\cite{elastica:Calugareanu:1961, topol:White:1969, elastica:Fuller:1971}, $Lk=Tw+Wr$. The geometric partitioning is amplified by the vanishingly small thickness of the nanoribbon that favors bends and twists relative to in-plane deformations, and forms the  basis for the paradigm that we employ to shape these nanoribbons. The approach is bioinspired in that it is also exploited to control the properties of natural filaments such as supercoiled DNA, $\alpha$-helices, elastomers, and textile fibers and their yarns~\cite{biofil:VinogradLaipis:1965, biofil:BolesCozzarelli:1990, biofil:StrickBensimon:1998, biofil:MorozNelson:1998, fil:AggeliBoden:2001, biofil:GhatakMahadevan:2005}, yet little is known about analogous conformations in these van Der Waals (vdW) nanoribbons. 

The rest of this article is organized as follows: we first present all-atom computations of twisted graphene nanoribbons subject to decreasing end displacements. The minimum energy conformations - consisting of scrolls, grafolds and supercoiled plectonemes - are systematically explore with varying degrees of initial twist and ribbon widths. Our results suggest a strong influence of the non-local van Der Waals interactions; we validate their effect by performing similar macro-scale experiments on thin magnetic and elastomeric ribbons.  A detailed theoretical analyses highlights the role of geometrical and physics parameters on the formation and stability of the conformations. We conclude with a brief discussion of the effect of these novel conformations on the nanoribbon properties, and potential applications. 
\section*{Results}
\begin{figure}
\centering
\includegraphics[width=\columnwidth]{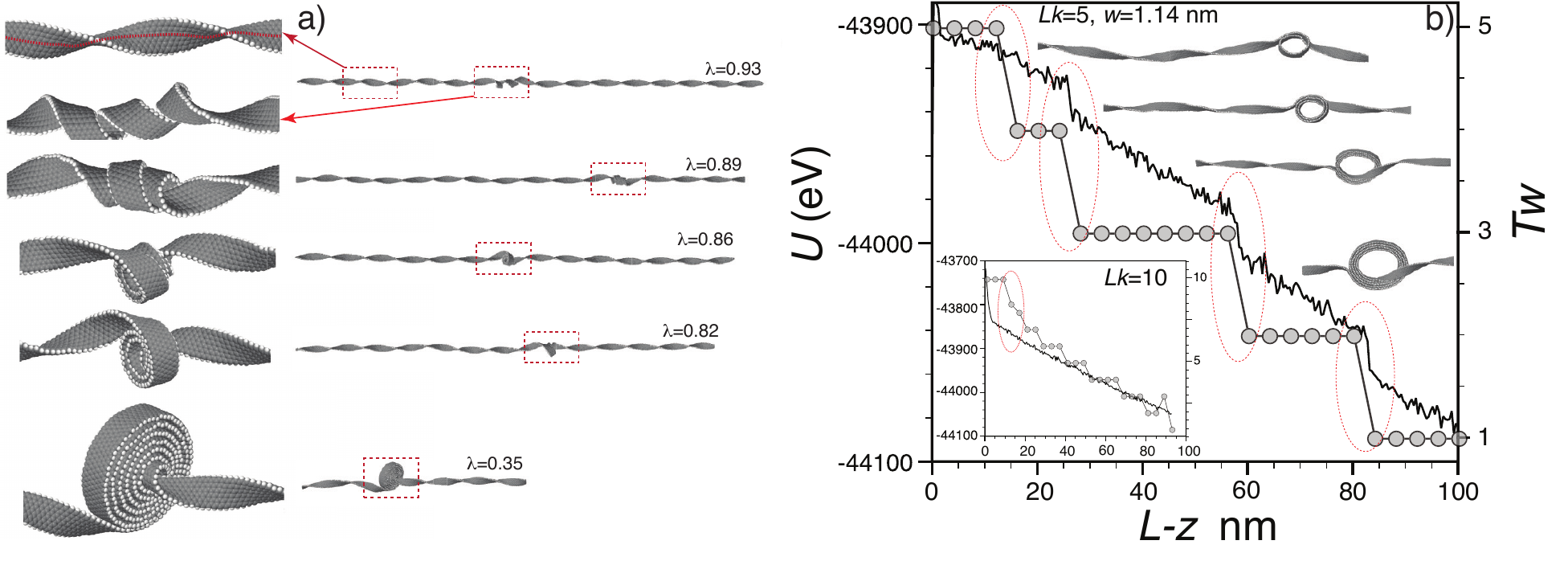}
\caption{
(a) Atomic configurations showing the formation and evolution of a scroll with decreasing $\lambda=z/L$ in a hydrogenated GNR of length $L=110.4$\,nm and width $w=1.136$\,nm, subject to a degree of supercoiling $Lk=10$. Carbon and edge hydrogen atoms are shaded gray and white, respectively. Expanded views of the boxed regions where the Writhe localizes are also shown. Some of the views are rotated to depict the details more clearly. The dotted red line in the $\lambda=0.93$ configuration traces the helical centerline of the ribbon. (b) The change in the interaction energy of the ribbon $U$ and the ribbon twist $Tw$ with end displacement $L-z$. The degree of supercoiling $Lk=5$ and $Lk=10$ in the main curve and inset, respectively.
\label{fig:fig2} 
}
\end{figure}

\noindent
{\bf Atomic-scale Computations}\\
The computations are performed on edge-hydrogenated zigzag GNRs as a function of ribbon supercoiling and end constraints. 
The stable conformations are extracted quasi-statically using molecular dynamics (MD) simulations (Methods).  Figure~\ref{fig:fig2}a and Supplementary Video 1 show the results of simulation of a GNR of length $L=110.4$\,nm and width $w=1.14$\,nm subject to a prescribed end rotation, $Lk=10$. The centerline of the initial $\lambda=1$ configuration is straight with a uniform twist density $\phi=2\pi\,(Lk/L)$.  
The end tension $T$ decreases with $\lambda$ and at a critical point the ribbon centerline destabilizes into a helix.
Each helical pitch $p$ contributes to a full rotation of the ribbon about its centerline (boxed region in Fig.~\ref{fig:fig2}a) as the large in-plane to bending stiffness ratio favors developable conformations with vanishingly small Gaussian curvature~\cite{book:Love:1944, elastica:RappaportRabin:2007}.





\subsection*{Scroll to fold transition} As $\lambda$ decreases, the helix radius increases and at $\lambda\approx0.93$, a localizing helical instability forms, grows in length, and beyond a critical size it 
packs itself into a dense scroll. 
The transition is apparent in Fig.~\ref{fig:fig2} for $\lambda=0.89$ and leads to the spontaneous formation of a multilayered scrolled at $\lambda=0.86$. The remainder of the ribbon visibly straightens following the scroll formation due to the concomitant increase in the axial tension. The scroll axis is inclined to the original ribbon axis and it is mobile along the ribbon length. Thereafter, the
scroll grows in size as the rest of the ribbon is reeled in, evident in the $\lambda=0.35$ conformation consisting of a 6-layer scroll bounded by twisted ribbon segments. 
In some cases the instabilities nucleate at multiple sites, in particular at larger widths, then rapidly diffuse along the ribbon length and try to coalesce. An example is shown in Supplementary Video 2  for $w=1.6$\,nm and $Lk=8.25$.

The inset in Fig.~\ref{fig:fig2}b shows the ribbon interaction energy $U$ and the Twist $Tw$ as a function of the imposed displacement $d=L-z$. The latter is extracted as the differential rotation of a material frame about the tangent $\hat{t}$ to the ribbon centerline (Supplementary Methods) and is simply one-half the number of local crossings of the two edges\cite{topol:Kauffman:2001, topol:DennisHannay:2005}. For $Lk=10$, the energy initially decreases rapidly as the ribbon destabilizes into a helix following the removal of the pre-stretch. Further decreasing $\lambda$ leads to a small jump when the scroll forms (encircled) and the energy then decreases steadily, albeit at a slower rate. $Tw$ evolves along expected lines; we see a spontaneous decrease following scroll nucleation, also evident in the configurations. 
Each subsequent transition decreases the twist by $\Delta Tw\approx1$ and the twist density decreases by $\Delta\phi\approx2\pi (1)/L_h$, where $L_h$ is the length of the helical phase. 
It is absorbed by the scroll and registers as a corresponding increase in $Wr$ (Supplementary Figure 2). This is immediately obvious as each layer or loop within the scroll contributes to non-local self-crossing of the centerline, the very definition of Writhe~\cite{elastica:Fuller:1971, topol:Kauffman:2001}. 

\begin{figure}
\centering
\includegraphics[width=\columnwidth]{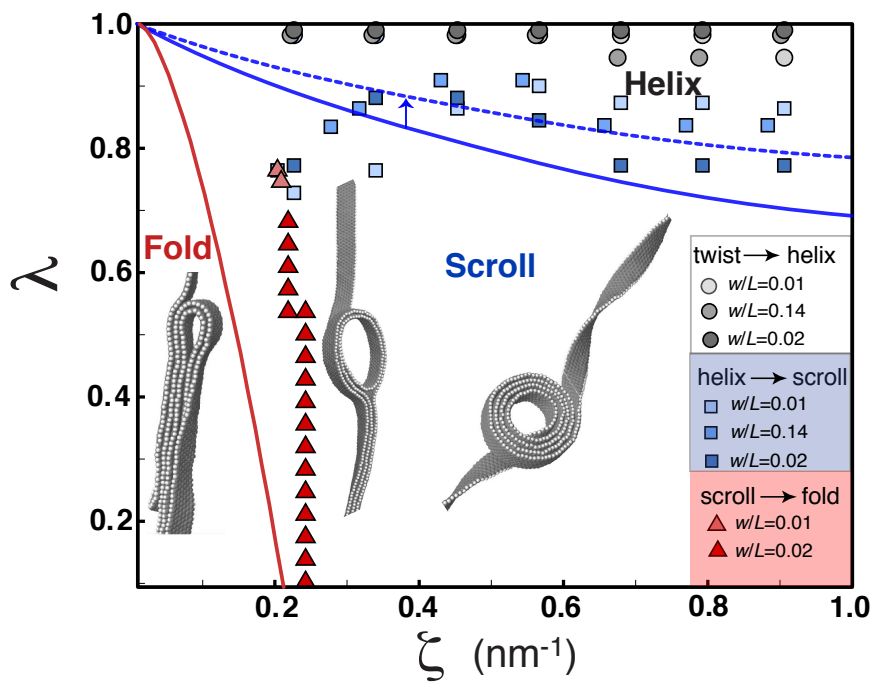}
\caption{The conformational phase diagram $\lambda$ vs $\zeta$.
The co-existence lines are the set of critical points ($\lambda$, $\zeta$) at which a new phase arises. The solid lines are the theoretically predicted critical curves. The gray, red and blue colors correspond to twist-helix, helix-scroll and scroll-fold transitions, respectively. The dotted blue line is prediction for the helix-scroll transition with a larger value of the interaction area fraction, $\alpha=2$.
\label{fig:fig3} 
}
\end{figure}
Figure~\ref{fig:fig2}b also shows the results for the same ribbon subject to a smaller Linking number, $Lk=5$. The reduced pre-stretch lowers the critical value of $\lambda$ for helix formation. The radius and the pitch length of the helix are larger and the initial energy decrease is therefore smaller. The nucleated scroll consists of a single loop that is partially double-layered with a larger inner radius. 
The transition leads to a significant decrease in both $U$ and $Tw$. Subsequently, they decrease sharply within narrow ranges of the end displacement (encircled). 
Each transition is preceded by fluctuations in the scroll size, aided by decreasing $\lambda$, as the scroll overcomes the barrier for incorporation of a new layer.
Initial scroll growth occurs at the expense of the helical phase, and at a critical size it forms a new layer by repacking itself into a smaller size with a concomitant decrease in the twist, $\Delta Tw\approx1$. The transition is spontaneous 
 since the ribbon ends terminating at the scroll must realign along the horizontal to maintain the force balance with the clamped ends.
Evidently, the energy barrier for scroll formation $\Delta U$ increases with decreasing $Lk$, and the degree of supercoiling serves as a driving force for the nucleation of the helical instability and its transformation into a scroll. Then, $Lk=10$ represents a relatively high driving force with almost continuous decrease in energy with $L-z$. Conversely, smaller supercoiling (e.~g.~$Lk=5$) enhances the role of fluctuations as the energy barrier is larger, and the formation and growth of the scroll is discontinuous.  

\subsection*{Scroll to fold transition} At even smaller $Lk$, we uncover yet another phase wherein the helical ribbon folds onto itself in a hairpin fashion. Occasionally, the multilayered segment is associated with a small twist. In some instances, we see tennis racquet like shapes consisting of coexisting folds capped by scrolls at one or both ends. 
We postpone the discussion of these shapes for now, 
but details of the evolution of these shapes are shown in Supplementary Videos 3 and 4 ($w=2.0$\,nm, $Lk=2.5$, and $w=1.1$\,nm, $Lk=3.75$, respectively).

\subsection*{Conformational phase diagram}
We develop a more complete understanding of the stable phases by performing simulations with varying $Lk$ and ribbon aspect ratios $w/L$ (Methods). The results are shown in Fig.~\ref{fig:fig3} as a conformational phase diagram, $\lambda$ vs. $\zeta=2\pi Lk/L$, the link density. Representative conformations are shown alongside.  
The initial twist is more stable at larger $\zeta$ due to the increasing pre-stretch, but in all cases small end displacements near $\lambda\approx1$ result in a spontaneous transition to a helix. 
The scrolled phase dominates 
 for smaller values of $\lambda$, as expected.  Close to the scroll-fold transition curve ($\zeta\approx0.25$ at small $\lambda$), we see co-existing scrolls and multilayered folds; an example conformation is shown alongside. High values of $\zeta$ do not lead to any new phases. The response is a bit different as the larger ribbon width and pre-stretch leads to scroll formation before the ribbon relaxes out the intrinsic twist and and the ribbon is stretched. The localized instability is a tightly wound helix that resembles an axially slit nanotube; an example is shown in Supplementary Figure 1. 

\subsection*{Plectoneme formation}
\begin{figure}
\includegraphics[width=0.9\columnwidth]{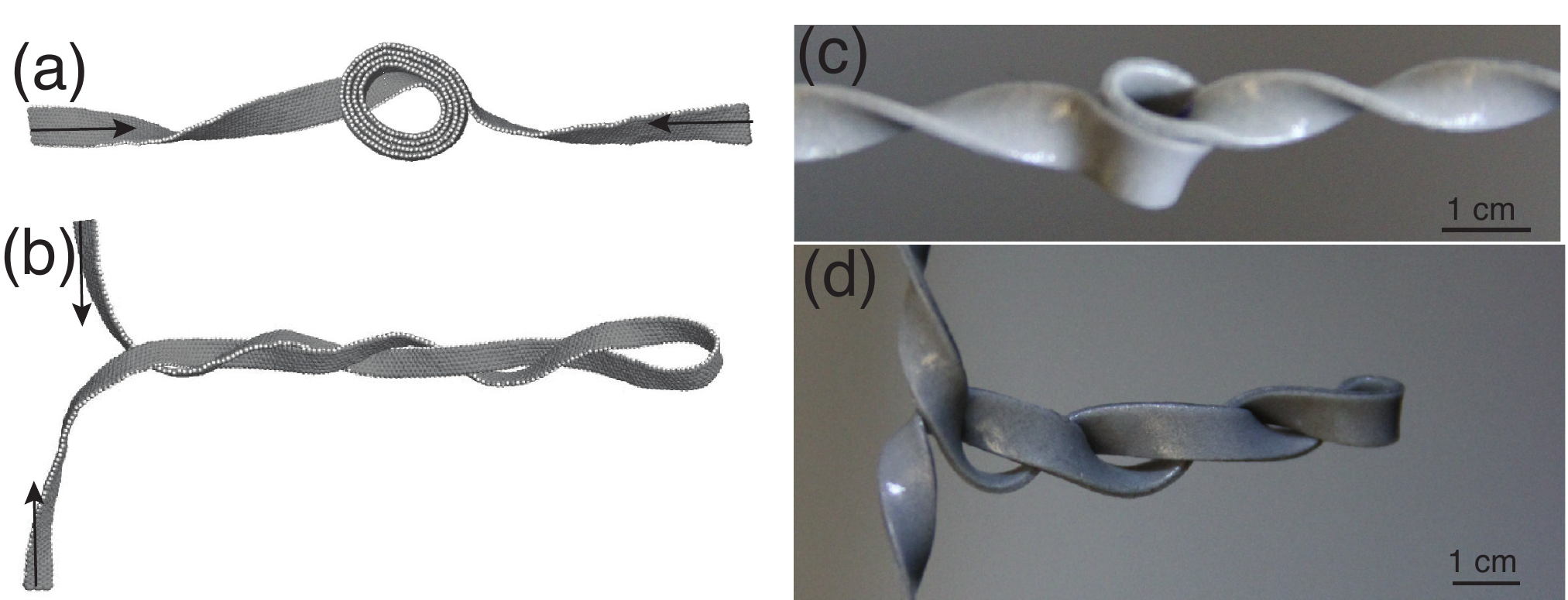}
\caption{(a-b) Atomic configurations of a GNR (a) with and (b) without long range vdW interactions. In both cases, the ribbon is subject to an end displacement of $\lambda=0.65$ (arrows). (c-d) Results of a macroscale experiments on (c) double-sided magnetic tape, and (d) simple elastic tape. The aspect ratio and the degree of supercoiling of the tapes are chosen to be the same as that of the GNRs, i.e. $w/L\approx0.01$ and $Lk=7.5$.
\label{fig:fig5}
}
\end{figure}

The dense scrolls and folds that we observe are rarely observed in soft ribbon-like assemblies due to the inherent self-avoidance in these solvated polymeric systems. There are exceptions, such as the formation of hairpin loops in folded $\beta$-sheet domains in proteins and in DNA/RNA which are stabilized by long-range interactions such as hydrogen bonding and specific base-pair interactions~\cite{book:KornbergBaker:1992}. 
 Clearly, the long-range vdW interactions have a decisive effect on the nature of the writhed conformations as they are comparable to the elastic energies associated with conformations i.e. bending and twist. 
As validation, we have repeated the computations by turning off the long-range vdW interactions. The direct comparison is shown in Figs.~\ref{fig:fig5}a~and~\ref{fig:fig5}b for a GNR of length $L=110$\,nm and width $w=2$\,nm, subject to supercoiling $Lk=7.5$ and an end displacement $\lambda=0.65$. The scroll formation is suppressed (Fig.~\ref{fig:fig5}a) and it now forms a classical plectoneme phase that grows with decreasing $\lambda$ (Fig.~\ref{fig:fig5}b, Supplementary Video 5). 

\section*{Macroscale Experiments}
We test if the behavior is universal by studying the effect of rigid loading conditions on macroscale elastic tapes with comparable aspect ratios and degree of supercoiling (Methods). In the case of flexible two-sided magnetic tapes ($0.076$\,cm thick, $w=0.5$\,cm, $L=50$\,cm), we see the nucleation of singly looped scrolls (Fig.~\ref{fig:fig5}c). Control experiments on non-magnetic tapes lead to the formation of classical plectonemes (Fig.~\ref{fig:fig5}d). The macroscale experiments serve as a useful validation, yet the dense phases are not routinely observed as the magnetic interactions are weak compared to the elastic energies. Gravity effects cannot be ignored as well. In contrast, the elastic energies stored in atomically thin ribbons are much smaller as they scale down with thickness and the effect of non-local interactions is therefore amplified. A simplified analysis on energetics of singly and doubly looped scrolls highlights the role of ratio of the bending stiffness of the GNR and the vdW interaction energy per unit area, $\sqrt{D/u_c}$, on stabilization of these dense phases (Supplementary Discussion).  Since the extrinsic effects are unavoidable in the macroscale tapes, we eschew their systematic study and rely on a simplified theoretical framework to further validate and analyze stability of the observed conformations.

\section*{Theoretical Analysis}
We make contact with the displacement-controlled response by determining the energy of formation of a stable scroll within a nanoribbon subject to an end displacement parameterized by $\lambda$, and then minimizing it with respect to geometric variables. We ignore the kinetics of the transient local bifurcation that precedes the nucleation of the scroll~\cite{elastica:Coyne:1990, elastica:HeijdenThompson:1998, elastica:ThompsonChampneysI:1996, elastica:ThompsonChampneysII:1996} and also the effect of the clamped ends. The remainder of the ribbon is assumed to be a helical space curve as observed in the computations. We limit our analysis to moderate supercoiling and narrow  inextensible ribbons, consistent with the large in-plane rigidities of these thin ribbons. 
Then, the energetics readily follows from the classical Love-Kirchoff framework for helical deformations of rods with appropriate modifications for the anisotropic cross-section of developable ribbons~\cite{elastica:HeijdenThompson:1998}.

The scroll removes length $L_s$ and link number $Lk_s$ from the helical phase.  Since the conformation must conserve the total length and Linking number, $l_h + l_s =1$ and $\rho_h+\rho_s=1$
where $l_h=L_h/L$ and $l_s=L_s/L$ are the normalized lengths, and $\rho_h=Lk_h/Lk$ and $\rho_s=Lk_s/Lk$ are the normalized Linking numbers associated with the two phases. 
Ignoring the more complex configurations at the interface between the two phases (see Fig.~\ref{fig:fig2}a), the scroll phase stores its contribution mostly as Writhe while the Twist is distributed over the helical phase. 
The change in the end distance $z={\lambda}L$ is absorbed by the helical phase. Inextensibility guarantees that the ribbon can only bend and therefore its deformed state can be completely determined from the conformation of its centerline. The helical space curve can be described in terms of its pitch $p=2\pi h=z/Lk_h$ and radius $r=\sqrt{L_h^2-z^2}/2{\pi}Lk_h$ which together define the pitch angle $\eta=h/r$ and the generalized curvature and torsion (Supplementary Equation 4).
We simplify the geometry of the scroll by considering the limit when the scroll radius is much larger than the equilibrium interlayer distance such that the curvature variations within the multilayers can be ignored. Then, the relevant variable is its average curvature $\kappa_s(l_s, \rho_s)=2\pi (Lk_s/L_s)=2{\pi} (Lk/L) (\rho_s/l_s)$.
The elastic energy stored in the helical phase follows from Sadowsky-W\"underlich functional for narrow inextensible ribbons~\cite{elastica:Sadowsky:1930, elastica:Wunderlich:1962}, 
\begin{align}
\label{eq:energyhelix}
U_h = \int_0^{L_h} \frac{D}{2}\kappa_h^2(1+\eta^2)^2 w\,ds.
\end{align}
The scroll is stabilized by a competition between bending and interaction energies,
\begin{align}
\label{eq:energyScroll}
U_s = \int_0^{L_s} \left(\frac{D}{2}\kappa_s^2-\alpha u_c\right)w\,ds,
\end{align}
where $u_c=1.5\,$eV/nm$^2$ is the interaction energy per unit area between parallel graphene sheets. The interaction area fraction $\alpha$ varies as the scroll grows: $\alpha\approx1/4$ within the single loop that nucleates at small link numbers ($Lk_s=1$) while the scrolls that nucleate at  larger Linking numbers are usually doubly looped ($Lk_s=2$) such that $\alpha\approx1\,\frac{1}{8}$ (Fig.~\ref{fig:fig2}a). As the scroll grows and becomes increasingly multilayered, $\alpha\rightarrow2$. Although an approximate expression for $\alpha(\rho_h, Lk)$ can be constructed (Supplementary Equation 10), for now we ignore the variation as part of the minimization presented below.  

Ignoring the interface regions, the ribbon energy density $(U_h+U_s)/(wL)$ can be expressed as a functional of the form $f(l_h, \rho_h, \lambda)$
which depends on $\beta=\zeta \sqrt{D/2\alpha u_c}$, a dimensionless parameter that captures the effect of link density and the competing effects of the bending stiffness and the interaction energy (Supplementary Equation 12). 
Minimizing $f$ with respect to the dimensionless twist $\rho_h$ and length $l_h$ yields their equilibrium values as a function of the end distance $\lambda$,
\begin{align}
\label{eq:helixscroll}
l_h (\lambda) = \lambda(\lambda+\beta),\quad
\rho_h(\lambda) = \frac{l_h^2 - \lambda^2}{l_h - \lambda^2}.
\end{align}
The solution to these equations partition the conformational space into three distinct stable regimes: helix, co-existing helix and scroll, and coexisting straight nanoribbon and scroll.  The corresponding critical curves are plotted in Supplementary Figure 3. Below, we use these relations to quantify the formation and growth of scrolls and their transition to folds. 

\subsection*{Scroll nucleation} 
For partially multilayered singly looped scrolls that nucleate at small $Lk$, the critical point can be expressed as $\rho_h^\ast=(Lk-1)/Lk$ with $\alpha\approx1/4$. Substituting in  Eq.~\ref{eq:helixscroll} yields the critical end displacement $\lambda^\ast_s$ and the complete solution is plotted in Fig.~\ref{fig:fig3} as the critical curve $\lambda^\ast_s$ vs. $\zeta^\ast_s$. The decrease in $\lambda^\ast_s(\zeta^\ast_s)$ is in agreement with trends in the simulations.  However, the quantitative agreement breaks down to some extent at large $Lk$; the critical point in the simulations is consistently higher due to several reasons: One, we ignore the effect of the clamped ends.
Two, the long-range interactions within the transient conformations that precede the scroll nucleation are ignored in the theoretical framework. Three, the radius of the scroll is approximated as a constant  and can lead to errors with increasing number of layers within the scroll. Four, the theory ignores the non-isometric deformations at the interface where the scroll transitions to a helix. The interfacial region becomes increasingly localized at large Linking numbers and can retain significant elastic energy.  Finally, as mentioned earlier, the criteria for the nucleation of the doubly looped scroll must be changed to $\rho_h^\ast=(Lk-2)/Lk$, with $\alpha\approx1\frac{1}{8}$. The modified theoretical curve is also plotted in Fig.~\ref{fig:fig3} (dashed line). The arrow in the plot indicates the approximate point at which the doubly looped scrolls become viable in the simulations. The comparison captures the effect of $\alpha$ on the critical point; $\lambda^\ast_s$ and $\zeta^\ast_s$ increase with $\alpha$ (Supplementary Figure 5) and the predictions are in quantitative agreement with the simulation results. 

These effects notwithstanding, the simulations and the theoretical analysis shows that $\lambda^\ast_s$ decreases non-linearly with $\zeta^\ast_s$. The origin of the decay can be understood by considering the  solution in the limit $Lk\gg 1$ (Supplementary Methods), $\lambda^\ast_s \approx {\sqrt{\beta^2+4}-\beta}/{2}$.
For $\beta\rightarrow0$ and sufficiently large $Lk$ the decay is almost linear,
$\lambda^\ast_s\approx1-\beta/2$. 
The analysis also yields the critical scroll size,
\begin{align}
\label{eq:criticalSize}
R^\ast_s=\frac{l_s^\ast}{2\pi Lk_s^\ast} \approx \sqrt{\frac{D}{2\alpha u_c}} \lambda^\ast_s,
\end{align}
i.e. it is proportional to the critical end distance and therefore decreases non-linearly with the corresponding link density $\zeta_s^\ast$.
The theoretical predictions are in agreement with the simulations, especially at small $Lk$. As an example, for $Lk=5$ the size of the singly looped scroll in the simulations is $R\approx0.3$\,nm (Fig.~\ref{fig:fig2}b) and that predicted by theory is $R\approx0.45$\,nm. The predicted size of the doubly looped scroll decreases to $R\approx0.4$\,nm for $Lk=10$ compared to $R\approx0.25$\,nm in the simulations. The computed sizes are consistently smaller, due to inaccuracies in the assumed interaction area fraction $\alpha$ at nucleation and related simplifying assumptions.

\subsection*{Scroll growth}
\begin{figure}
\centering
\includegraphics[width=0.75\columnwidth]{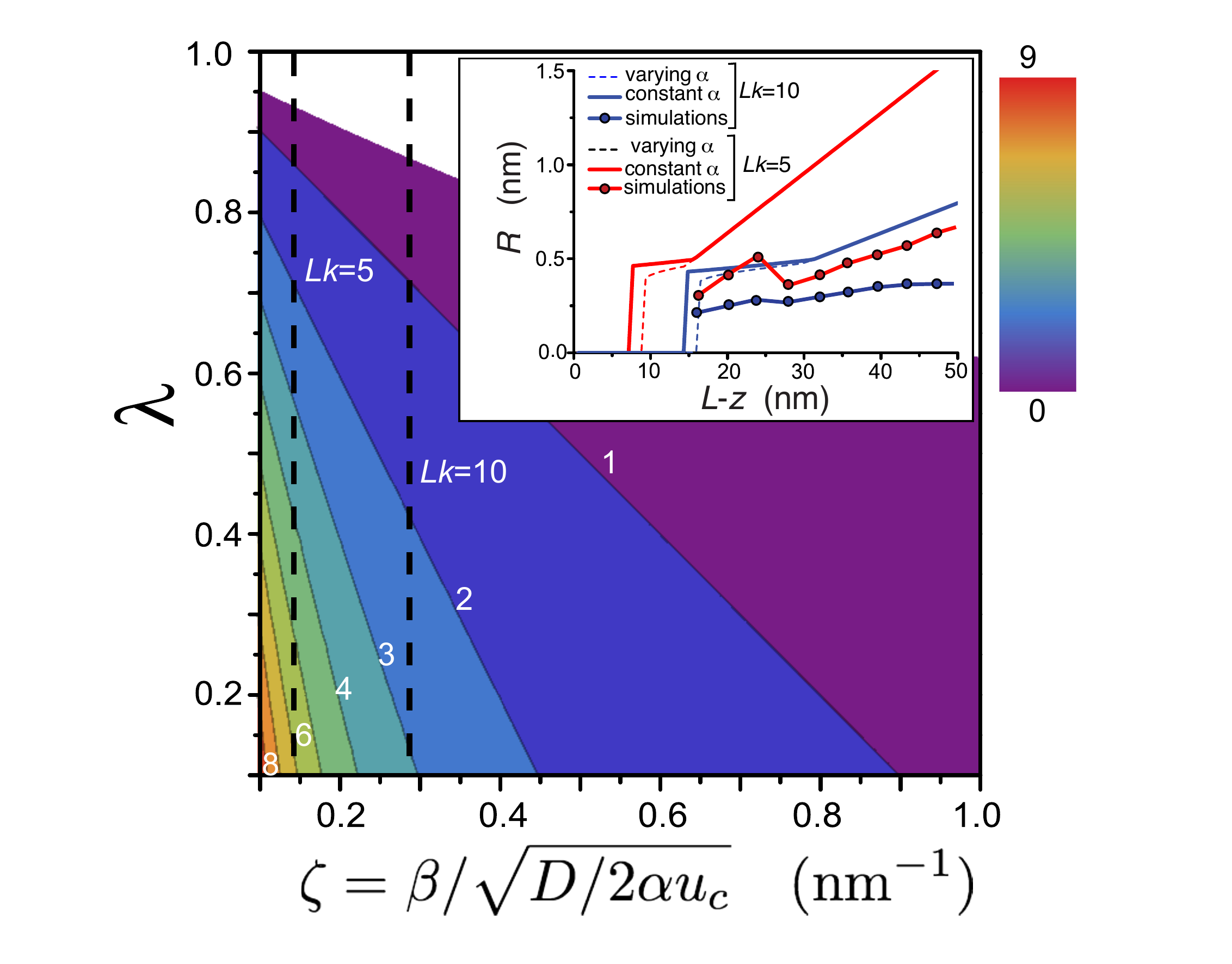}
\caption{\label{fig:fig6} Dimensionless contour plot of predicted scroll size $R(\lambda, \zeta)$ in a GNR of length $L=110$\,nm and width $w=1.14$\,nm. (inset) The evolution of the scroll size with end-distance $L-z$ for $Lk=5$ (red solid line) and $Lk=10$ (blue solid line) with constant interaction area fraction $\alpha=2$. Theoretical predictions for varying $\alpha$ (dashed lines) and the size evolution extracted from the simulations (symbols) are also plotted. See text and Supplementary Methods for details.}
\end{figure}

The scroll size in the simulations increases with decreasing $\lambda$, evident in Fig.~\ref{fig:fig2}. The behavior is consistent with Eq.~\ref{eq:helixscroll},
\begin{align}
\label{eq:growth}
R(\lambda, \beta) = \sqrt{\frac{D}{2\alpha u_c}} \left(\frac{1}{\lambda + \beta}\right).
\end{align}
and the dependence is plotted in Fig.~\ref{fig:fig6} for the GNRs geometries studied here. Quantitative agreement with the simulations is handicapped as $\alpha$ is assumed to be a constant. In order to fully capture this effect, we have repeated the energy minimization with varying $\alpha(\rho_h, Lk)$ (Supplementary Figure 4) and the predicted size evolution $R(\lambda)$ is plotted in the inset in Fig.~\ref{fig:fig6} for $Lk=5$ and $Lk=10$. Comparison with plots for constant $\alpha$ 
(vertical dashed lines) indicate that the rapid increase in $\alpha$ clearly tempers the initial scroll growth. 

Figure~\ref{fig:fig6} also shows the size evolution in the simulations. The nucleated sizes are smaller yet the initial growth rate is in excellent agreement with the theory. Past a critical size, the scroll size begins to increase discontinuously with the addition of a new layer, slowing the growth rate. The behavior is more pronounced at lower $Lk$ and can be clearly seen in the $Lk=5$ curve. This is a result of mechanical equilibrium along the horizontal that aligns the ribbon tangents at the interface regions abutting the scroll along the clamped ends. The preferred orientation also minimizes the distortion at the interface regions. As $\lambda$ decreases, the constraint forces the scrolls to grow by increasing the scroll length $L_s$; the decreasing end distance directly feeds the scroll by increasing its size. Past the critical point, it becomes energetically favorable for the scroll to repack by increasing the numbers of layers and therefore $\alpha$ (Fig.~\ref{fig:fig2}b). This entails sliding between the layers in the scroll. Since the interlayer interactions are weak in these vdW materials, the energy dissipation is negligible. The addition of each new layer is spontaneous once the existing scroll overcomes the energy barrier.  


\subsection*{Scroll to fold transition} 
Low link densities lead to large scroll sizes that are unstable due to the small bending stiffness of the scrolled segment. Then, past another critical size, it becomes energetically favorable for the scroll to collapse into bi-/multi-layered folds. The transition is again spontaneous as unfolding requires larger end forces.  Usually, the folding transition occurs well after the remainder of the ribbon has eliminated all of its twist, i.e. $L_h=0$. In this regime, $\lambda<1-\beta$, the end distance $L_h\propto z$, and the scroll radius evolves as (Supplementary Methods)
\begin{align}
\label{eq:scrollGrowthFold}
R_s(\lambda, \beta) = \sqrt{\frac{D}{2\alpha u_c}} \left( \frac{1-\lambda}{\beta} \right).
\end{align}
The phenomenon is similar to the self-collapse of nanotubes where the enhanced interaction energy between the collapsed layers stabilizes the large curvatures at the folded ends~\cite{cnt:TangGlassmaker:2005}. The critical radius of a scroll that undergoes energetically favored self-collapse is
\begin{align}
\label{eq:fold}
R_{f}^\ast \approx 2.124\sqrt{\frac{D^\prime}{u_c}}, 
\end{align}
where $D^\prime$ is the effective bending rigidity of the multilayered scroll and varies with the number of layers. The size is much bigger than the interlayer distance $d_0$ such that $D^{\prime}/u_c\gg d_0$. 
The fold transition in the simulations occurs usually for multilayered scrolls (see Supplementary Video 4, $Lk=3.25$) and $\alpha=2$ is a reasonable approximation for the scroll radius (Eq.~\ref{eq:scrollGrowthFold}). Additionally, since the resistance to interlayer sliding is small, the bending rigidity is simply the independent contribution of each layer and can be approximated as $D^{\prime}\approx DLk_s(=DLk)$. Then, equating Eqs.~\ref{eq:scrollGrowthFold}~and~\ref{eq:fold}, $\lambda_f^\ast \approx 1- 4.248\, \beta \sqrt{Lk}$.
The curve, plotted in Fig.~\ref{fig:fig3}, is in good agreement with the simulation results, especially so for large end displacements where the critical point is almost independent of $\lambda$. Then, the critical link density varies inversely with ribbon length, $\zeta^\ast_f\propto (D L/u_c)^{-1/3}$. The extent of the folded region is proportional to the critical size and it grows with decreasing $\lambda$ following nucleation; the behavior is similar to the scroll growth analyzed earlier.

\noindent
\section*{Discussion and Conclusions}  
Our results demonstrate a simple strategy for controlling the size and number of layers in these packed phases via geometric end constraints, thereby enabling continuous on-demand modification of their properties. The ability to prescribe their shapes offers a novel route for tuning the ribbon properties, and is therefore of importance for their deployment as active elements in emerging nanoelectronic devices, electromechanical systems and nanocomposites.  In particular, the higher level of control can aid the development of a novel class of non-linear nanoelectronic and NEMS devices - actuators,  resonators switches - based on these vdW materials. The paradigm also applies to nanoribbons in other material systems, notably carbon nanotubes and their bundles\cite{cntr:LiangUpmanyu:2005b, asit:LiangUpmanyuMahajan:2008, nt:SomuUpmanyu:2010, cnt:HahmWangUpmanyuJung:2012}, and nanowires and nanoribbons of  polar crystals such as ZnO and GaN where the non-local interactions are considerably stronger due to the presence of surface charges\cite{nw:PanWang:2001, nw:ChenHaataja:2011}. The interplay with end-constraints highlighted here can be employed to engineer a far richer set of conformations with their own unique set of properties - there is plenty of room at the bottom in controlling the conformations of these nanoribbons.

\section*{\large Methods}
{\bf Atomic-scale simulations}: The GNRs chosen for this study have zigzag edges. Carbon atoms at the unreconstructed zigzag edges with one missing $sp^2$ bond are passivated with hydrogen atoms\cite{gnr:LeeCho:2009}. The simulations are performed for fixed ribbon length $L=110.4$\,nm and varying widths, $w=1.136$\,nm, $1.562$\,nm, and $1.988$\,nm. The AIREBO framework is used to describe the carbon-carbon bonded interactions in the GNRs\cite{intpot:Brenner:2002}.  The vdW interactions are based on the classical $6-12$ potential between graphene elements\cite{cntr:Girifalco:2000}. The potential reproduces the near-equilibrium properties of graphene; since this study is limited to soft conformations, the empirical framework is adequate for this study. The degree of supercoiling is prescribed by twisting the ribbon uniformly along its length with a twist density $\phi=2\pi \,Lk/L$. The ends are then clamped such that the end distance is equal to the  contour length of the untwisted ribbon.
The ribbon is relaxed  using canonical MD at a $T=300^\circ$K (velocity Verlet integrator, time step 1\,fs, Nos{\'e}-Hoover thermostat\cite{book:AllenTildesley:1989, md:Plimpton:1995}). 

\noindent
{\bf Ribbon conformation}: The effect of $\lambda$ on the conformations is explored by decreasing the end distance quasi-statically in decrements of $2\%$ and the energy is locally minimized using the MD algorithm. The ribbon vectors $\hat{u}(s)$ are the generators of the developable ribbon surface (see Fig.~\ref{fig:fig1}). A subset of these vectors terminate at the passivating hydrogen atoms at the edges, and they are monitored to dynamically generate the ribbon shape and extract $Tw$ and $Wr$ (Supplementary Methods).

\noindent
{\bf Phase diagram}: The co-existence lines represent the combination of parameters associated with the first observation of a stable new phase. The critical point for twist to helix transition is based on destabilization of the ribbon centerline; it develops a finite curvature with a well-defined pitch length smaller than the ribbon contour length. The helix to scroll transition follows from observations of stable loops. Singly looped scrolls are commonly observed at small $Lk$ ($Wr=1$, Fig.~\ref{fig:fig2}b) while larger supercoilings result in doubly looped scrolls ($Wr=2$, Fig.~\ref{fig:fig2}a). Formation of both scrolls and folds results in discontinuous changes in the interaction energy $U$ and aids in identifying the corresponding critical points.
 
\noindent
{\bf Macroscale experiments}:
The magnetic tapes were cut out from double sided magnetic sheets (McMaster-Carr, $\sim 4$\,kPa pull strength). The ends are gripped and rotated using clamps. The end distance is decreased at a rate $\sim10^{-3}$\,s$^{-1}$. The control experiments are performed on non-adhesive marking tape (3M).

\pagebreak


\noindent

\clearpage

\noindent

{\bf Acknowledgements}: The computations were performed on {\it st}AMP and Discovery supercomputing resources at Northeastern University. The study is supported by seed grant from Northeastern University (AS), and National Science Foundation DMR CMMT (1106214, MU and AS) and DMREF CHE (14348424, MU) Programs.

\noindent

{\bf Author Contributions}: MU conceived the study. AS performed the computations. AS and MU devised and performed the macro scale experiments. HW and MU developed the theoretical frameworks. All authors analyzed the results and wrote the paper.

\pagebreak
\section*{\bf \centering Supplementary Material}

\subsection*{Supplementary Discussion}
A simplified energetic analysis presented here sheds light on the role of the weak van der Waal interactions on the stability of the scrolls and folds. Following Ref.[9], consider the initial stages of the formation of a scroll of size $R$ in a nanoribbon with $Lk=1$. An end displacement of $L-z=2\pi R$ removes the twist by nucleating a partially bilayered singly looped scroll as seen in Fig.~4c. The change in energy is
\begin{align}
\label{eq:eq1}
\Delta U 
&=\frac{D}{2}\frac{2\pi R}{R^2} + 2\pi R (T-\alpha u_cw) - 2\pi |M|.
\end{align}
The first term is the bending energy, the second term is the work done by the average end tension $T$, the third term represents the energy gain due to the long-range interactions within the bilayered portion, and the last term is the energy gain as the Twist $Tw$ is converted to Writhe $Wr$. Increasing the initial rotation to $Lk=2$ leads to a two-stage response where the end displacement $2\pi R$ first leads to a singly looped scroll with the residual twist $Tw=1$. The energetics can be described by a modified version of Supplementary Equation~\ref{eq:eq1} that also includes the energy stored in the twist, which scales quadratically with the twist density $\phi=Tw/\lambda L$. For $L-z\approx4\pi R$, the loop grows by a layer and forms a doubly looped scroll, and in the process removes the twist completely in the remainder of the ribbon. The additional energy change can be again described by Supplementary Equation~\ref{eq:eq1}. A limitation of this analysis is that since the couple ($M$, $T$) varies with $L-z$, further analytical progress requires a load-controlled experiment (fixed $M$ or $T$) where $Lk$ is now allowed to adjust to the imposed dead load. For example, for constant $M=C\phi$ with $C$ the torsional rigidity, minimizing $\Delta U$ yields the critical scroll size $R^\ast=\sqrt{D/2T^\prime}$ with a modified adjusted tension, $T^\prime=T-\alpha u_c w$. The corresponding energy change $\Delta U = \sqrt{D/2T^\prime} - |M|$.  The effect of supercoiling is embedded in the end couple as it increases with $Lk$. Although the analysis does not make direct contact with the rigid loading employed in simulations, it reveals the importance of the ratio $D/u_c$ in stabilizing these conformations.

\section*{Supplementary Methods and Equations}
\subsection*{Additional simulation details}

{\bf Stability of initial twisted configuration}:
The initial end distance $z$ can be in principle larger than the ribbon contour length $L$ ($\lambda>1$). The resultant pre-stretch is higher. However, it does not lead to qualitatively new behavior. For $\lambda=1$, the phase diagram shows that the twisted structure is usually stable. We have confirmed this for large initial twists $Lk\ge10$ by performing simulations in excess of $10$ \,ns 
with $\lambda=1$. 
Note that at large supercoiling, the initial ribbon conformation is no longer isometric due to the large pre-stretch. An example of such a conformation is shown in Supplementary Figure~\ref{fig:Supp1}. As $\lambda$ is decreased, the pre-stretch is gradually eliminated and the conformation becomes isometric (also shown in the figure).
\begin{figure*}[htb] 
\centering
     \includegraphics[width=0.7\columnwidth]{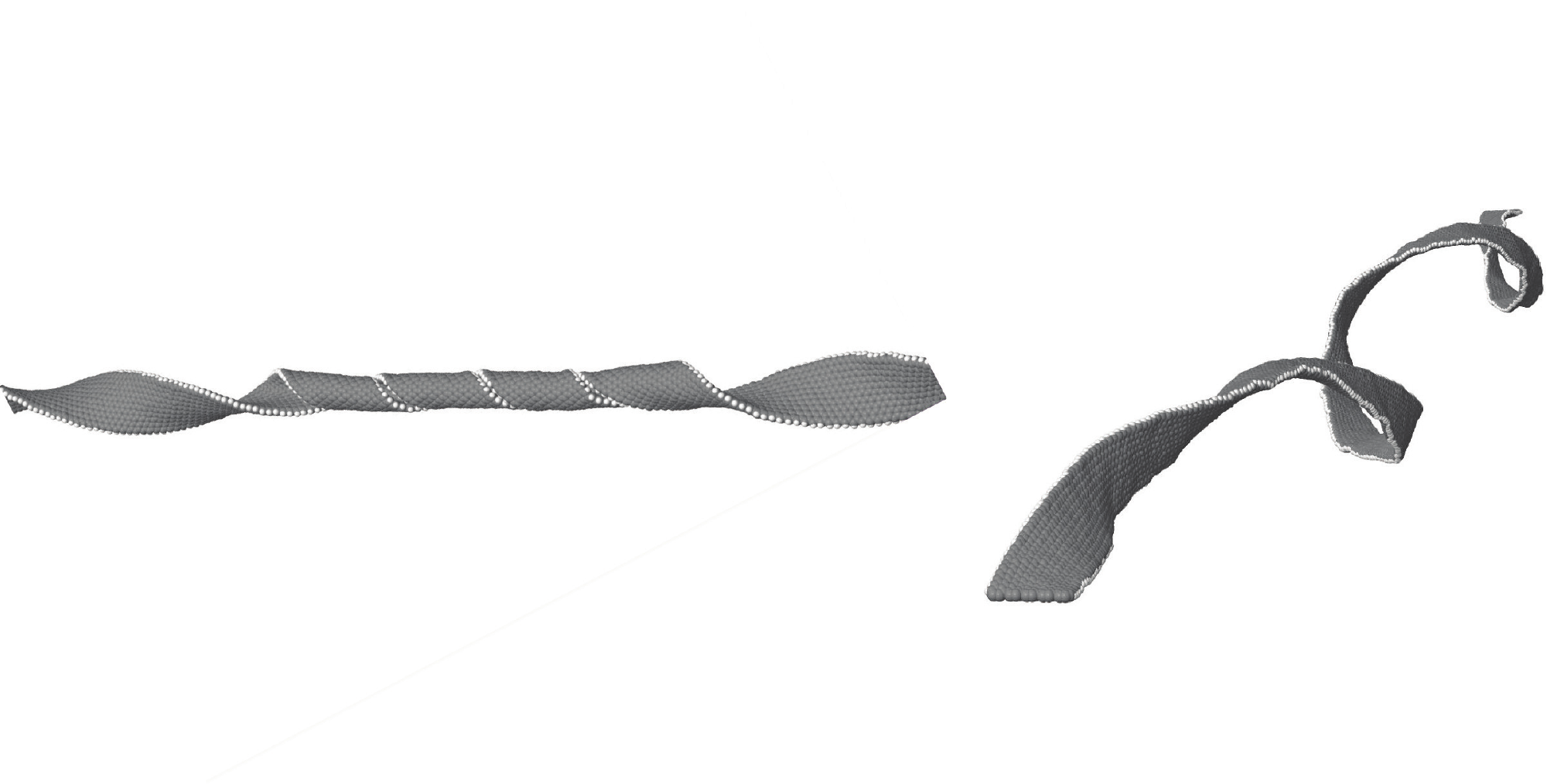}
   \caption{\label{fig:Supp1} (left) Twist to helical transition at a large Linking number, $Lk=17$ and $\zeta\approx1$\,{nm}$^{-1}$. The pre-stretch due to the twist-stretch coupling in the ribbon results in a stretched and tightly wound helix. (right) For $\lambda<1$, the ribbon releases the stretch and adopts an isometric conformation.}
\end{figure*}

{\bf Low temperature simulations}:
To quantify the effect of thermal fluctuations on the critical points, we have performed a series of MD simulation at $10$\,K for a nanoribbon with $L=110.4$\,nm, $w=1.36$\,nm and $Lk=10$. The critical points for twist to helix transition and  scroll formation ($\lambda_s^\ast$, $\zeta_s^\ast$) are unchanged. The temperature independence of the results is not surprising as the ribbon length is much smaller than its persistence length. However, the transitions are smoother and the scrolls do not diffuse along the ribbon length. 

\subsection*{Extraction and analysis of ribbon conformations}
{\bf Ribbon centerline and edges}: The initial relaxed, untwisted ribbon geometry is used to link each vector with its terminal pair of hydrogen atoms. Thereafter, they are generated dynamically by monitoring the coordinates of the linked hydrogen atoms. The average of the coordinates of the pairs of hydrogen atoms yields a point on the centerline. The entire centerline is generated by  interpolating a space curve through these points. The ribbon edges are constructed similarly by interpolating through the coordinates of the terminal hydrogen atoms.

\noindent
{\bf Twist and Writhe}: The ribbon centerline and the edges are used to extract $Tw$ and $Wr$ stored in the ribbon. $Tw$ is the rotation of the ribbon vector $\hat{u}$ about the ribbon tangent $\hat{t}$,
\begin{align}
Tw = \frac{1}{2\pi}\int_0^L \left[\left(\hat{t}\times \hat{u}\right)\cdot \frac{d\hat{u}}{ds}\right] ds=\frac{1}{2\pi}\int_0^L \Omega_t ds.
\end{align}
It is extracted by counting the number of times the pair of hydrogen atoms exchange their position about the ribbon centerline, we ignore the fractional twist. $Wr$ is a more involved, non-local double integral. For continuous deviations from a reference curve which we take to be the horizontal that runs through the clamped ends $\hat{z}$, the curve simplifies to a local quantity [11, 7]
\begin{align}
Wr \approx \frac{1}{2\pi}\int_0^L \frac{1}{1+\hat{t}\cdot\hat{z}}\left[\left(\hat{t}\times \hat{z}\right)\cdot \frac{d\hat{t}}{ds}\right] ds.
\end{align}
It is calculated continuously. Note that there are no antipodal points ($\hat{t}\cdot\hat{z}=-1$) in the scroll or the fold as their axis is always inclined to $\hat{z}$.

The results for a ribbon of width $w=1.14$\,nm, length $L=110.4$\,nm and $Lk=10$ are shown in Supplementary Figure~\ref{fig:Supp2}. There are fluctuations in the quantities due to the finite temperature, yet the Linking number extracted as the sum $Tw+Wr$  is invariant, as expected.
\begin{figure*}[htb] 
\centering
     \includegraphics[width=0.7\columnwidth]{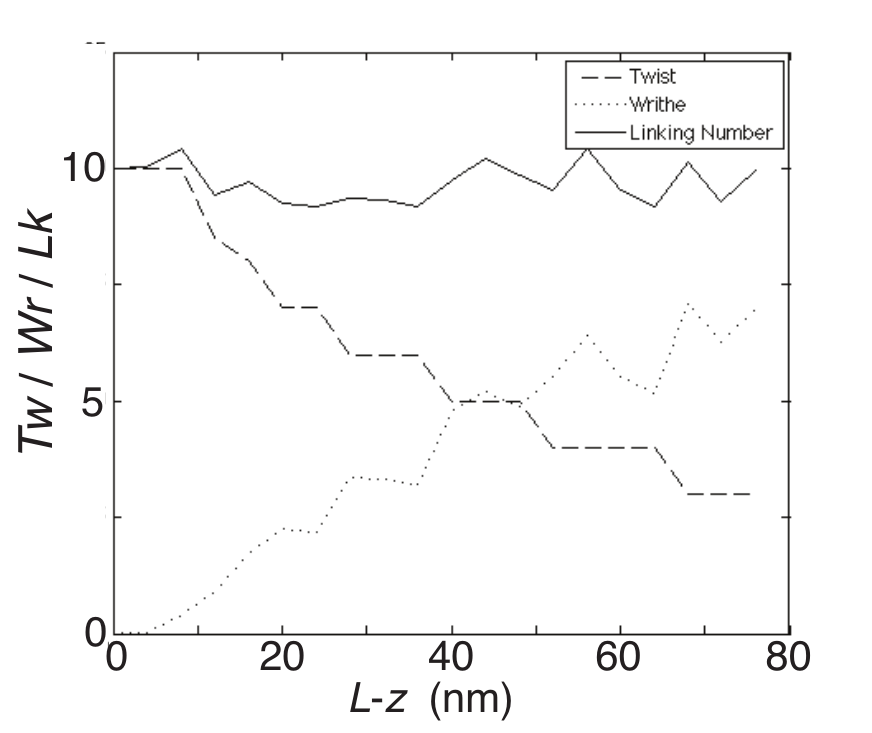}
   \caption{ \label{fig:Supp2} The evolution of the extracted $Tw$ and $Wr$ for a zigzag GNR with $Lk=10$. The fractional $Tw$ is ignored. The Linking number is plotted as the sum $Tw+Wr$.}
\end{figure*}

\subsection*{Theoretical analyses}
{\bf Rigidly loaded nanoribbon}: The nanoribbon centerline is modeled as an inextensible but deformable space curve due to the large stretching stiffness relative to bending and twist energies. We restrict the analysis to rigid loading conditions with a prescribed end distance $\lambda<1$ and fixed Linking number $Lk$. The critical point for helix-scroll transition follows from an energy balance between the two conformations. The energy of the helix is the sum of bending and twisting energies associated with the space curve traced by its centerline. The scroll regions are stabilized by vdW interactions between the layers, and energy associated with bending the layers. Below, we analyze a simple but illustrative example where the curvature and torsion of the ribbon centerline are uniform in the helical and scroll regions. We also ignore corrections due to discontinuities and highly localized deformations at the scroll-helix interface. 

\noindent
{\bf Geometry}: 
The nucleation of the scroll removes a length $L_s$ and Linking number $Lk_s$ from the helix.  
The total length and linking number should be conserved,
\begin{align}
\label{Constrain}
&l_h + l_t + l_s=1\nonumber\\
& \rho_h+\rho_t+\rho_s=1.
\end{align}
Here, the normalized lengths $l_h=L_h/L$ and $l_s=L_s/L$, and the normalized Linking numbers $\rho_h=Lk_h/Lk$, and $\rho_s=Lk_s/Lk$, with $L_h$ and $L_s$ the lengths of helical and scrolled regions, and $Lk_h$ and $Lk_s$ the corresponding Linking numbers. For the sake of completeness, the expression is generalized to also include the normalized length and Linking number associated with intrinsic twist in the nanoribbon, $l_t=L_t/L$ and $\rho_t=Lk_t/Lk$. The centerline of the helix,  constrained to an end distance $z=L(\lambda - l_t)$, results in a pitch $2{\pi}h=z/Lk_h$ and helical radius $r_h=\sqrt{(L_h)^2-z^2}/2{\pi}Lk_h$. Then, the generalized curvature $\kappa_h$, torsion $\tau_h$ and radius $r_h$ can be expressed as,
\begin{align}
\label{Helix}
& \frac{1}{r_h}=\frac{2{\pi}Lk}{L}\frac{\rho_h}{\sqrt{(\l_h)^2-(\lambda-l_t)^2}}\nonumber\\
& \kappa_h=\frac{r}{r^2+h^2}=\frac{2{\pi}Lk}{L}\frac{\rho_h\sqrt{(\l_h)^2-({\lambda-l_t})^2}}{(\l_h)^2}\\
& \tau_h=\frac{h}{r^2+h^2}=\frac{2{\pi}Lk}{L}\frac{\rho_h(\lambda-l_t)}{(\l_h)^2}.\nonumber
\end{align}
The energetically relevant term for scroll is its average curvature $\kappa_s=1/r_s$,
\begin{align}
\label{Scroll}
\kappa_s=\frac{2{\pi}Lk_s}{L_s}=\frac{2{\pi}Lk}{L}\frac{\rho_s}{l_s}.
\end{align}

\noindent
{\bf Energetics}:
The Gaussian curvature of an inextensible ribbon is invariant under isometric deformations of the surface. For a developable ribbon without spontaneous curvature, the Frenet frame (FF) and the material frame (MF) are coincident and the internal twist  $\tau_t=0$ ($l_t=\rho_t=0$)~[18]. In the limit of small ribbon widths $w\ll L$ as is the case here, the free energy of helical region follows from the Sadowsky functional~[25]
\begin{align}
\label{EnergyH}
U_h=\frac{1}{2}D(\frac{1}{r_h})^2L_h w=\frac{1}{2}D\frac{(\kappa_h^2+\tau_h^2)^2}{\kappa_h^2}L_hw,
\end{align}
where the out-of-plane bending stiffness $D=Et^3/12(1-\nu^2)$.
Supplementary Equation~\ref{EnergyH} expressed in terms of the pitch angle $\eta=\tau_h/\kappa_h$ yields Eq.~1 in the main text. 

The free energy of the scroll $U_s$ is the sum of bending energy $U_s^b$ and torsional energy $U_s^{int}$,
\begin{align}
\label{EnergyS}
& U_s=U_s^b+U_s^{int}=[\frac{1}{2}D\kappa_s^2-{\alpha}u_c]L_sw,
\end{align}
where $\alpha u_c$ is the interaction energy per unit area of the scroll as mentioned in the text.
Combining Supplementary Equations~\ref{Constrain}-\ref{EnergyS} with $l_t=0$ and $\rho_t=0$ yields the total free energy of formation of a scroll within a helical nanoribbon,
\begin{align}
\label{Energy}
 U=U_h+U_s &=\left\{\frac{2{\pi}^2 Lk^2 D}{L^2}\left(\frac{\l_h\rho_h^2}{\l_h^2-\lambda^2}\right) + \frac{2{\pi}^2 Lk^2 D}{L^2}\left(\frac{\rho_s^2}{l_s}\right)-{\alpha}u_cl_s \right\}Lw\nonumber\\
& = \left\{\frac{2{\pi}^2 Lk^2 D}{L^2}\left[\frac{\l_h\rho_h^2}{\l_h^2-\lambda^2}
 + \frac{(1-\rho_h)^2}{1-\l_h}\right]-{\alpha u_c}(1-\l_h) \right\}Lw.
\end{align}
The interaction area increases with scroll size and this effect can be captured by allowing $\alpha$ to vary. Singly looped scrolls nucleate such that only a small part of the loop is double layered and  interacting. We take this limit to be $\alpha\approx1/4$. The other limit is that of a multiple layered scroll,~i.~e. $\alpha\approx2$. The Linking number stored in the scroll is the measure of the number of layers. Taken together with the two extreme cases, the fractional interaction area $\alpha(\rho_h, Lk)$ takes the form
\begin{align}
\label{Alpha}
\alpha\approx2(\frac{Lk_s-1}{Lk_s})+\frac{1}{4Lk_s}=\frac{8(1-\rho_h)-7/Lk}{4(1-\rho_h)}.
\end{align}
Substituting in Supplementary Equation~\ref{Energy}, we get,
\begin{align}
\label{EnergyV}
& U=\{\beta^2[\frac{\l_h\rho_h^2}{\l_h^2-\lambda^2}
 + \frac{(1-\rho_h)^2}{1-\l_h}]-\frac{4(1-\rho_h)-7/2Lk}{4(1-\rho_h)}(1-\l_h)\}\alpha u_cLw,
\end{align}
with $\beta=\sqrt{D/2\alpha u_c} (2\pi Lk/L)=\zeta\sqrt{D/2\alpha u_c}$. 
The limit $Lk\gg1$ yields the energy functional referenced in the main text,
\begin{equation}
\label{EnergyC}
\begin{aligned}
& U/Lw=\{\beta^2[\frac{\l_h\rho_h^2}{\l_h^2-\lambda^2} + \frac{(1-\rho_h)^2}{1-\l_h}]-(1-\l_h)\}\alpha u_c= f(l_h, \rho_h, \lambda).
\end{aligned}
\end{equation}

\noindent
{\bf Stability Diagram and Scroll Size Evolution}

\noindent
{\bf Constant $\alpha$ ($Lk\gg1$)}:
Minimization of Supplementary Equation~\ref{EnergyC} gives us the critical values of $\rho_h$ and $\l_h$  as a function of $\lambda$ and $\beta$. Minimizing first with respect to $\rho_h$,
\begin{align}
\label{Min1}
& \frac{{\partial}f}{{\partial}\rho_h}=2\beta^2[\frac{\l_h\rho_h}{\l_h^2-\lambda^2}-\frac{(1-\rho_h)}{1-\l_h}]\alpha u_c=0
\quad \rightarrow \quad \rho_h=\frac{\l_h^2-\lambda^2}{\l_h-\lambda^2},
\end{align}
with the solution of interest in the range $0<\rho_h<1$ for $\lambda\leq\l_h\leq1$. Substituting Supplementary Equation~\ref{Min1} into the energy functional,
\begin{align}
f=\frac{\l_h^2+\lambda^2+\l_h(\beta^2-1-\lambda^2)}{\l_h-\lambda^2}\nonumber,
\end{align}
and minimizing with respect to length of the helical phase $l_h$, we get
\begin{align}
\label{Min2}
\frac{{\partial}f}{{\partial}\l_h}=0 \quad \rightarrow \quad \l_h=\lambda(\lambda\pm\beta).
\end{align}
$\lambda(\lambda-\beta)<\lambda$ is not permissible and the solution of interest lies in the range $\lambda<\lambda(\lambda+\beta)<1$. Combining Supplementary Equations~\ref{Min1} and~\ref{Min2}, we arrive at the three stable conformational regimes:
\begin{align}
\label{Transition1}
&\begin{cases} 
\l_h=1 \\ 
\rho_h=1 
\end{cases}
& \lambda>\frac{\sqrt{\beta^2+4}-\beta}{2} & \quad \rightarrow \quad {\rm helix}\\
\label{Transition2}
&\begin{cases} 
 \l_h=\lambda(\lambda+\beta) \\ 
\rho_h=\lambda \left[(\lambda+\beta)^2-1 \right] /\beta
\end{cases}
& 1-\beta<\lambda<\frac{\sqrt{\beta^2+4}-\beta}{2}, & \quad \rightarrow \quad {\rm helix+scroll}\\
\label{Transition3}
&\begin{cases} 
\l_h=\lambda \\ 
\rho_h=0 
\end{cases}
& \lambda<1-\beta, &\quad  \rightarrow \quad  {\rm straight + scroll}.
\end{align}
The results are plotted in Supplementary Figure~\ref{PhaseDiagram} as a stability diagram $\lambda$ vs. $\beta$.
\begin{figure}[ht]
\begin{center}
\centerline{\includegraphics[width=0.6\textwidth]{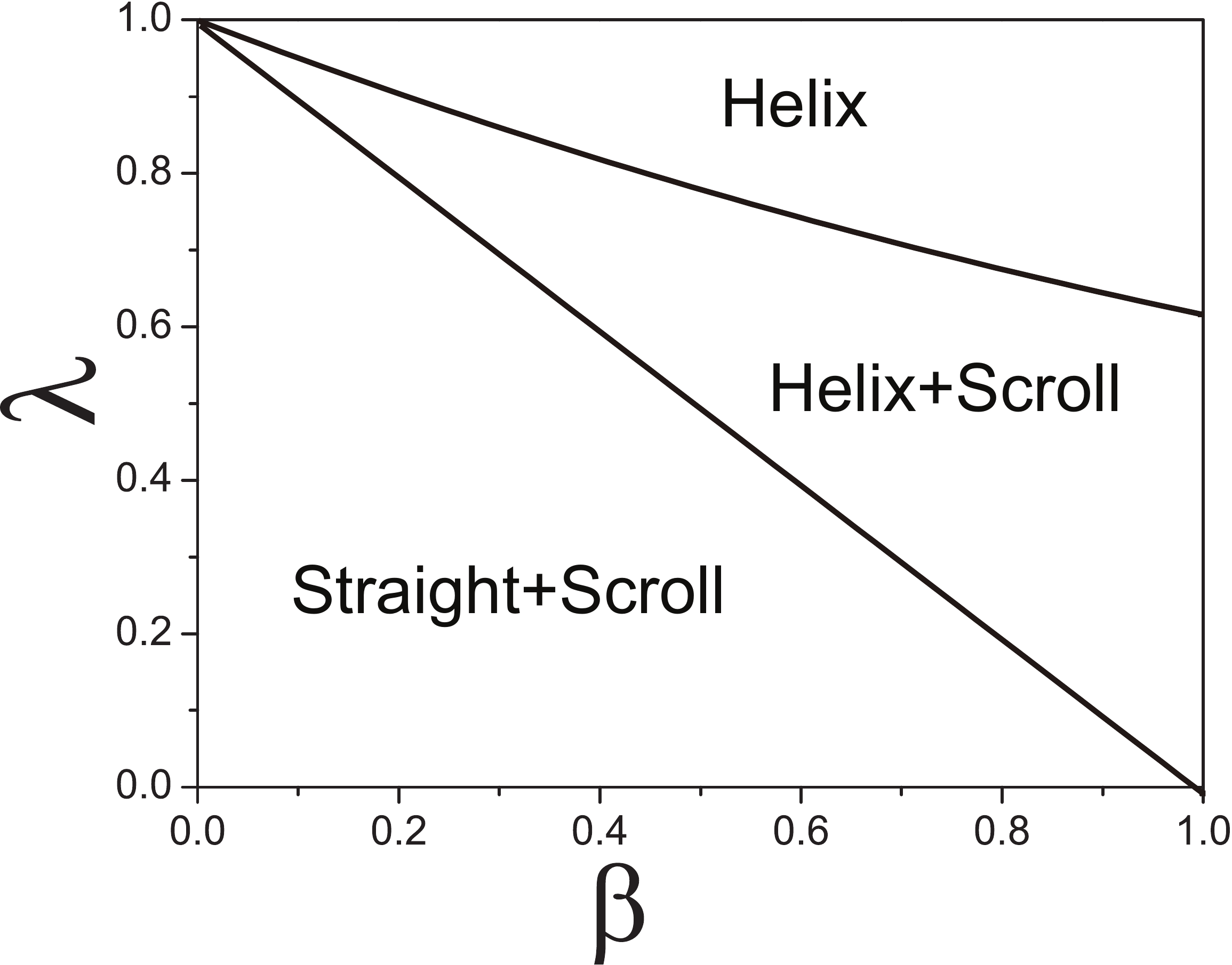}}
\caption{\label{PhaseDiagram} Stability diagram $\lambda$ vs. $\beta=\zeta\sqrt{\frac{D}{2\alpha u_c}}$ following energy minimization of a GNR.}
\end{center}
\end{figure}
The energy minimization also yields the equilibrium scroll radius, 
\[
R_s=\frac{Ll_s}{2\pi Lk \rho_s}=\frac{l_s}{\rho_s}\frac{1}{\beta}\sqrt{\frac{D}{2\alpha u_c}}
\]
in each of the three regions. Using Supplementary Equations~\ref{Transition1}-\ref{Transition3},
\begin{align}
\label{Radius}
&\begin{cases} 
\l_s=0 \\ 
\rho_s=0\\
R_s=0
\end{cases}
& \lambda>\frac{\sqrt{\beta^2+4}-\beta}{2} & \quad \rightarrow \quad {\rm helix}\\
&\begin{cases} 
l_s=1-\lambda(\lambda+\beta) \\ 
\rho_s=1-\lambda[(\lambda+\beta)^2-1]/\beta\\
R_s=\sqrt{\frac{D}{2\alpha u_c}} \;1/(\beta+\lambda)
\end{cases}
& 1-\beta<\lambda<\frac{\sqrt{\beta^2+4}-\beta}{2} & \quad \rightarrow \quad {\rm helix+scroll}\\
&\begin{cases} 
l_s=1-\lambda \\ 
 \rho_s=1\\
R_s=\sqrt{\frac{D}{2\alpha u_c}} \left(1-\lambda\right)/\beta
\end{cases}
& \lambda<1-\beta & \quad \rightarrow \quad {\rm straight+scroll}.
\end{align}
The variation is plotted in Fig.~5 for a zigzag GNR.


\noindent
{\bf Varying $\alpha(\rho_h, Lk)$}: We minimize Supplementary Equations~\ref{Alpha} and~\ref{EnergyV} numerically using {\it Mathematica}. The comparison with the analytical approximation for $L=110.4$\,nm, $w=1.136$\,nm and  $Lk=5$ and $10$ ($\beta=0.145, 0.290$) is plotted in Supplementary Figure~\ref{TheorySimu}.
\begin{figure*}[h!tp]
\begin{center}
\centerline{\includegraphics[width=\columnwidth]{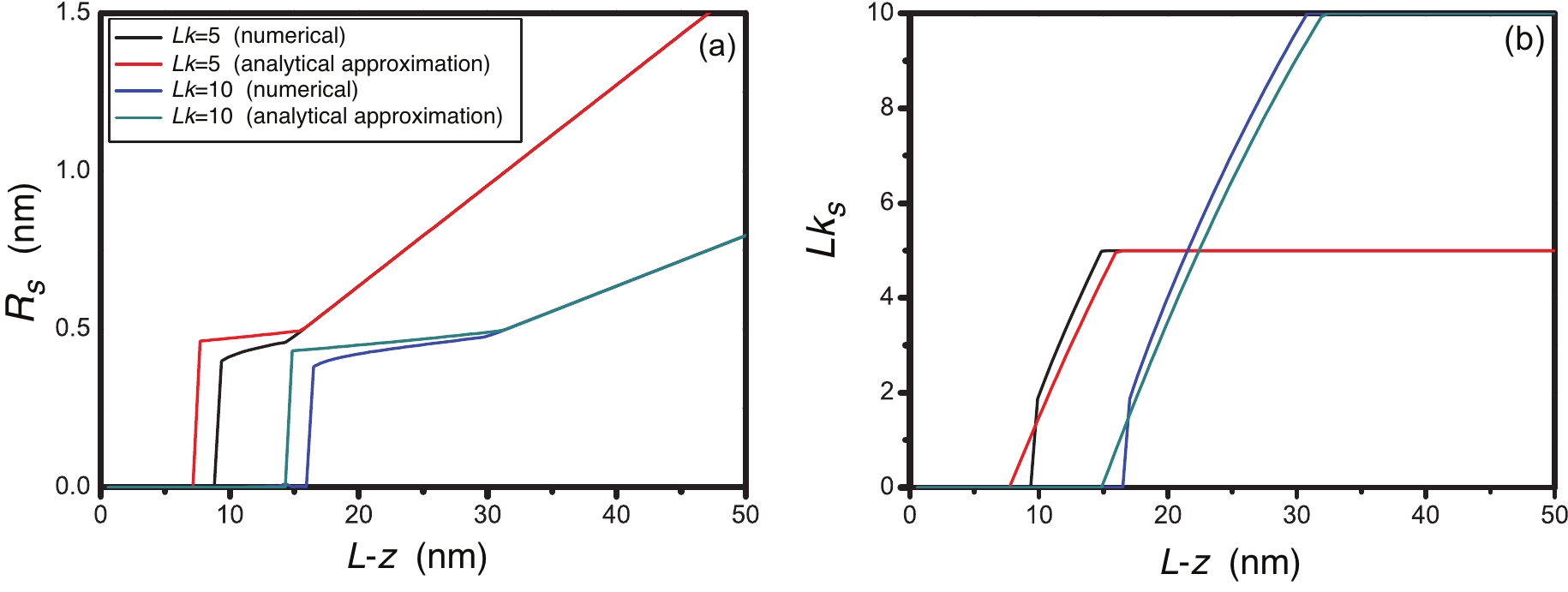}}
\caption{\label{TheorySimu} Evolution of (a) scroll radius $R_s$ and (b) and stored Linking number $Lk_s$ with end displacement $L-z$ for  $Lk=5$ and $Lk=10$. Both the analytical approximation and the the complete numerical solution are plotted.}
\end{center}
\end{figure*}

\noindent
{\bf Scroll Nucleation}
The critical values $\l_h^\ast$, $\rho_h^\ast$ and $R_s^\ast$ are obtained as function of $\beta$ and $Lk$ by solving Supplementary Equation~18,
\begin{align}
\label{eq:S17}
& l_h^\ast=\lambda^\ast(\lambda^\ast+\beta)\\
\label{eq:S18}
& \rho_h^\ast=1-\frac{{Lk_s^\ast}}{Lk}=\frac{{l_h^\ast}^2-{\lambda^\ast}^2}{l_h^\ast-{\lambda^\ast}^2}\\
\label{Radius}
&R_s^\ast=\sqrt{\frac{D}{2\alpha u_c}} \frac{1}{\beta+\lambda^\ast}
\end{align}
Substituting Supplementary Equation~\ref{eq:S17} in Supplementary Equation~\ref{eq:S18}, we get
\begin{align}
\label{Nucleation}
\lambda^\ast_s=\frac{1}{6}(-4\beta+2^{\frac{4}{3}}(3+\beta^2)\gamma^{-\frac{1}{3}}+2^{\frac{2}{3}}\gamma^{\frac{1}{3}}),
\end{align}
where 
\begin{align}
\gamma= & \left[9-27({Lk_s^\ast}/{Lk}) \right]\beta+2\beta^3\nonumber\\
&+3\sqrt{3}\sqrt{-4+[27({Lk_s^\ast}/{Lk})^2-18(Lk_s^\ast/{Lk})-1]\beta^2-4(Lk_s^\ast/{Lk})\beta^4}.\nonumber
\end{align}

The stable scroll after nucleation is singly looped ($Lk_s^\ast=1$) or doubly looped ($Lk_s^\ast=2$).  Defining $Lk_s^\ast=1$ as point at which the scroll nucleates and taking the large Linking number limit $Lk\gg Lk_s^\ast$, Supplementary Equations~\ref{Nucleation} and~\ref{Radius} can be further simplified as 
\label{Nucleation1}
\begin{align}
\label{simplifiedSoln1}
& \lambda^\ast_s=\frac{1}{2}\left(\sqrt{\beta^2+4}-\beta\right)\\
\label{simplifiedSoln2}
& R_s^\ast=\frac{1}{2} \sqrt{\frac{D}{2\alpha u_c}}\left(\sqrt{\beta^2+4}-\beta\right)
\end{align}
Furthermore, in the limit $\beta\rightarrow0$, we get
\begin{align}
\label{Nucleation2}
& \lambda^\ast_s=1-\frac{\beta}{2}\\
& R_s^\ast=\sqrt{\frac{D}{2\alpha_0 u_c}}\left(1-\frac{\beta}{2}\right)
\end{align}
The full solution in Supplementary Equation~\ref{Nucleation} and the simplified expressions (Supplementary Equations~\ref{simplifiedSoln1} and \ref{simplifiedSoln2}) are plotted in Supplementary Figure~\ref{Critical} for $L=110.4\,$nm and $w=1.136$\,nm, and $Lk_s^\ast=1$. Evidently, for small $\beta$ as in our simulations, the difference is negligible.
\begin{figure*}[ht]
\begin{center}
\centerline{\includegraphics[width=\textwidth]{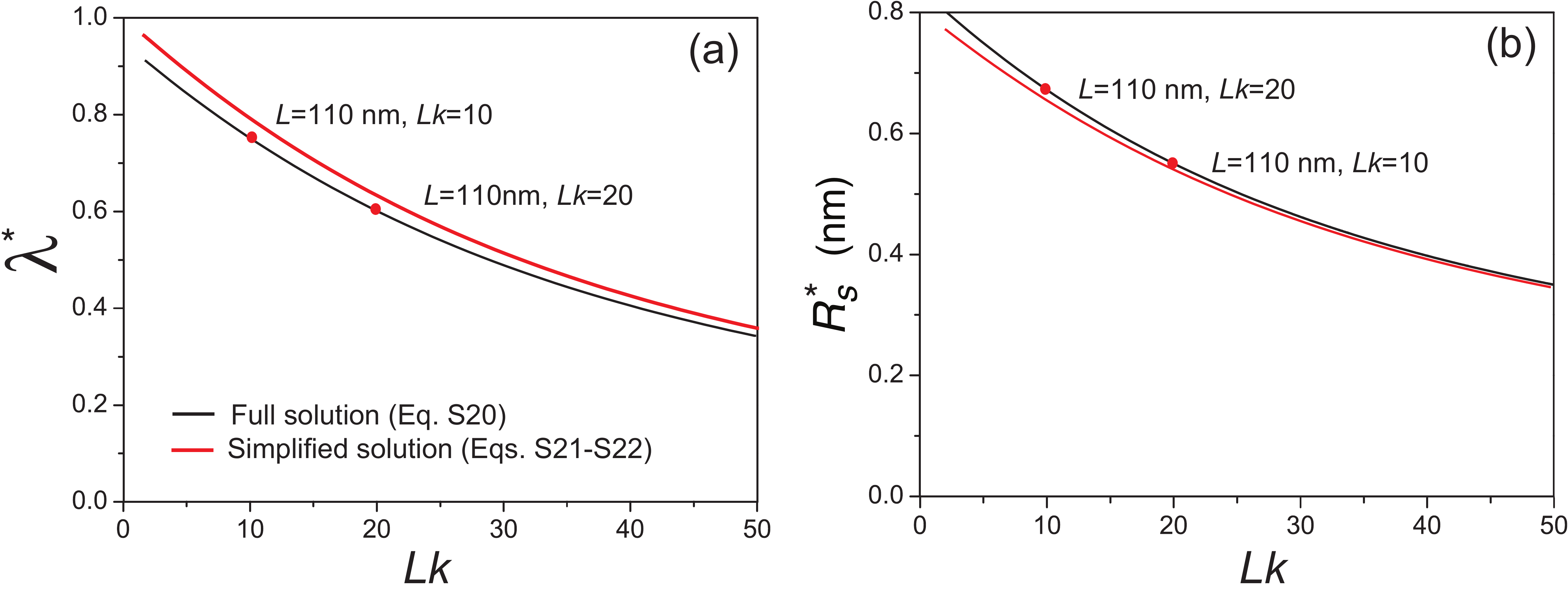}}
\caption{\label{Critical} (a) Critical end distance $\lambda^\ast_s$ and (b) scroll  radius $R_s^\ast$ as function of the Linking number $Lk$.}
\end{center}
\end{figure*}



\noindent
{\bf Self-collapse: Scroll-fold transition}
The critical scroll radius for energetically favored self-collapse is a low-dimensional analogue of the collapse of large radii nanotubes, and we use a similar approach to analyze this transition.
Following Tang et al.~[27], the critical radius can be expressed as
\begin{align}
\label{Tang}
R_f^\ast=A+\sqrt{A^2-B^2}
\end{align}
with
\[
A=\frac{4.076}{\pi} \left(\sqrt{\frac{D^\prime}{\alpha u_c}-0.13d_0}\right) \quad {\rm and} \quad B=\sqrt{\frac{D^\prime}{\alpha u_c}}.
\]
Here, the layer distance $d_0=0.34$\,nm and $\alpha=1$. Note that the original solution is for a single-walled nanotube and therefore corresponds to a singly-looped scroll. For, multi-layers, the enhanced bending stiffness can be approximated as the sum of the contribution of the individual layers that slide with respect to each other, $D^\prime (Lk_s)=Lk_sD$. Further, in the limit ${D}/{u_c}\gg d_0$ we can ignore the effect of the finite inter-layer thickness. Then, Supplementary Equation~\ref{Tang} simplifies to,
\begin{equation}
\label{Tang2}
R_f^\ast=2.124\sqrt{\frac{DLk}{u_c}}
\end{equation}


The precursor scroll that collapses is assumed to be multilayered, and Supplementary Equation~\ref{Radius} with $\alpha=2$ is the corresponding critical size of the scroll, and we arrive at the criteria for self-collapse,
\begin{align}
\label{Flat0}
& \frac{1}{2}\sqrt{\frac{D}{u_c}}\frac{(1-\lambda)}{\beta}>2.124\sqrt{\frac{DLk_s}{u_c}}, {\rm with}\\
\label{Flat1}
& Lk_s=Lk.
\end{align}
The second condition (Supplementary Equation~\ref{Flat1})  simply states that the remainder of the nanoribbon is untwisted, i.e. the scroll collapses only after it has grown to a point that it removes the twist completely. The fold transition captured by Supplementary Equation~\ref{Flat0} can be expressed as function of $\zeta=2\pi Lk/L$,
\begin{align}
\label{Flat2}
\lambda^\ast_f < 1-4.248\beta\sqrt{Lk} & =1-4.248\pi\frac{(Lk)^{3/2}}{L}\sqrt{\frac{D}{u_c}}\nonumber\\
& =1-0.847\sqrt{\frac{DL}{u_c}}\zeta^{3/2}
\end{align}
The critical curve ($\lambda^\ast_f, \zeta_f^\ast$) for zigzag GNRs with $D=1.5$\,eV/nm$^2$, $u_c=1.536$\,eV, $L=110.4$\,nm and $w=1.36\,$nm is shown in Supplementary Figure~\ref{Flat} and is also plotted in Fig. 3 in the main text.
\begin{figure*}[ht]
\begin{center}
\centerline{\includegraphics[width=0.5\textwidth]{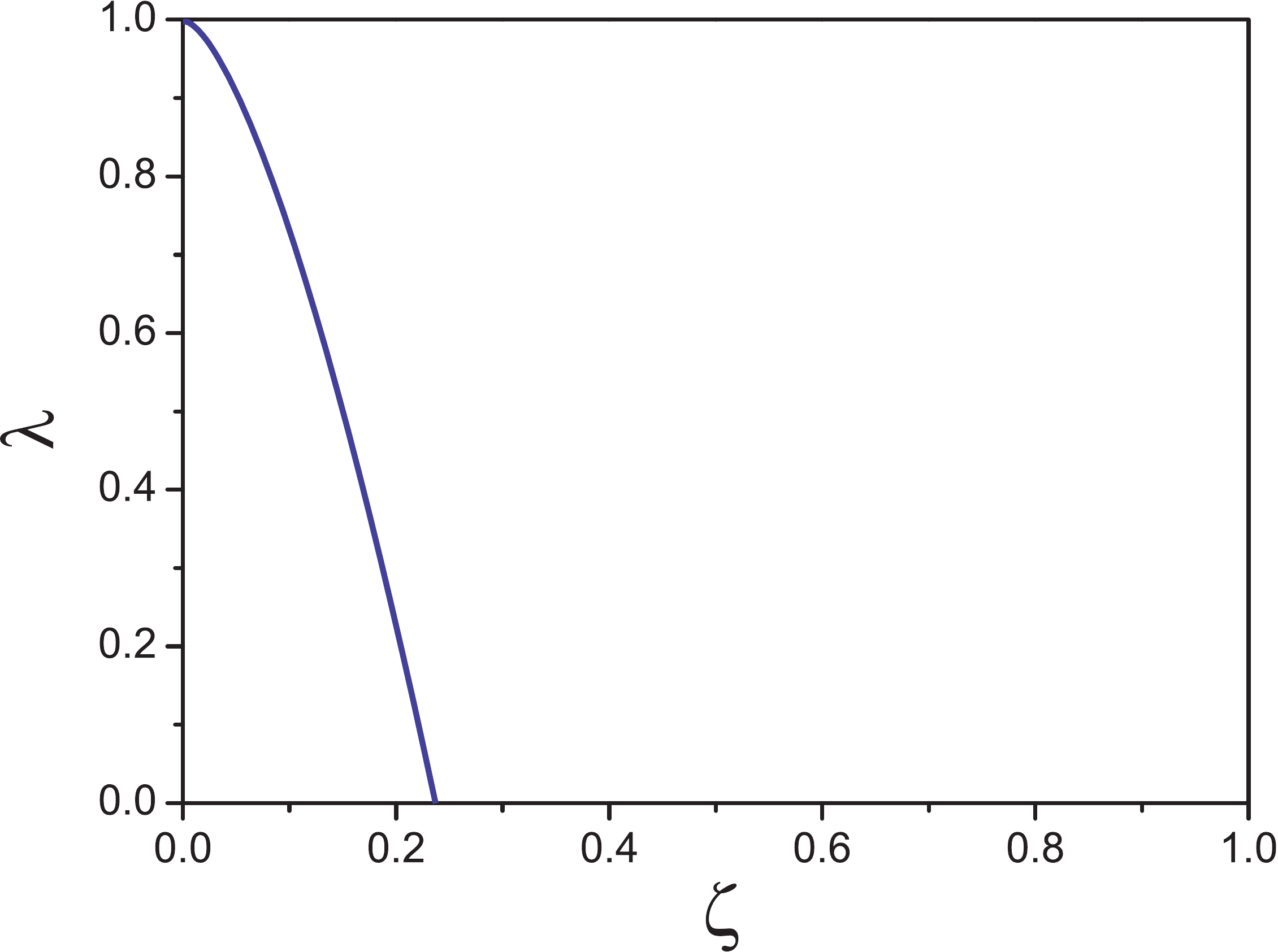}}
\caption{\label{Flat} The scroll/flat transition curve, plotted as the critical stretch $\lambda_f^\ast$ versus the critical link density $\zeta_f^\ast$.}
\end{center}
\end{figure*}

\end{document}